\documentclass[twocolumn,floatfix,amsmath,amssymb]{revtex4}
\usepackage{graphicx} \usepackage{dcolumn} \usepackage{color}

\newcommand{\be}{\begin{equation}} \newcommand{\ee}{\end{equation}}

\def\eg{{\it e.g.}\ }  
  
\def\rhs{{\it r.h.s.}\ }

 \def\els{Els\"asser variables\ }

\def \pmbtext#1{\leavevmode \setbox0\hbox{#1}
     \kern-0,2pt \copy0 \kern-\wd0 \kern0,4pt \copy0 \kern-\wd0
     \kern-0,2pt \raise0,3pt \box0 }

 \newcommand{\bb}{\mathbf{b}}
\newcommand{\bA}{\mathbf{A}} 

\newcommand{\alp}{\alpha}

\newcommand{\bj}{\mathbf{j}} 
\newcommand{\bom}{\mbox{\boldmath $\omega$}}

\newcommand{\bompm}{\mbox{\boldmath $\omega^{\pm}$} }

\newcommand{\bv}{\mathbf{v}}

\newcommand{\bzpm}{\mathbf{z^{\pm}}} \newcommand{\bzmp}{\mathbf{z^{\mp}}}

\begin{document}
\title{Small scale structures in three-dimensional magnetohydrodynamic turbulence}

\author{P.D. Mininni$^1$, A.G. Pouquet$^1$ and D.C. Montgomery$^2$}
\affiliation{$^1$ NCAR, P.O. Box 3000, Boulder, Colorado 80307 \\
 $^2$ Dept. of Physics and Astronomy, Dartmouth College, Hanover, 
NH 03755}
\date{\today}
\begin{abstract}
We investigate using direct numerical simulations with grids up to 
$1536^3$ points,  the rate at which small scales develop in a decaying 
three-dimensional MHD flow both for deterministic and random initial conditions.
Parallel current and vorticity sheets form at the same spatial locations, 
and further destabilize and fold or roll-up after an initial exponential 
phase. At 
high Reynolds numbers, a self-similar evolution of the current and vorticity maxima  is found, in which they grow as a cubic power of time; the flow then reaches a finite 
dissipation rate independent of Reynolds number.
\end{abstract}
\maketitle

Magnetic fields are ubiquitous in the cosmos and play an 
important dynamical role, as in the solar wind,
stars or the interstellar medium.  Such flows have large Reynolds 
numbers and thus nonlinear mode coupling  leads to the formation of  
strong intermittent structures. 
It has been observed that such extreme events in magnetohydrodynamics (MHD) 
are more intense than for fluids; for example, wings of Probability 
Distribution Functions of field gradients are wider and one observes 
a stronger 
departure from purely self-similar linear scaling with the order of the 
anomalous exponents of structure functions \cite{carbone}.  
Since Reynolds numbers are high but 
finite, viscosity and magnetic resistivity play a role, 
tearing mode instabilities develop and reconnection takes place. 
The question then becomes at what rate does dissipation occur, as 
the Reynolds number increases? What is the origin of these structures, 
and how fast are they formed?

This is a long-standing problem in astrophysics, \eg in the context of reconnection events in the magnetopause, or of heating of solar and stellar corona. 
In such fluids, many other phenomena may have to be taken into account, 
such as finite compressibility and ionization, leading to a more complex Ohm's law 
with e.g. a Hall current 
or radiative or gravitational processes to name a few.
Many aspects of the two-dimensional (2D) case are understood, but the three-dimensional (3D) turbulent case 
remains more obscure.
Pioneering works \cite{green}
show that the topology of the reconnecting region,
more complex than in 2D, can lead to varied behavior.

The criterion for discriminating between a singular and a regular behavior 
in the absence of magnetic fields follows the seminal work by Beale, Kato 
and Majda (hereafter BKM) \cite{BKM} where, for a singularity to develop in the 
Euler case, the time integral of the supremum of the vorticity must grow as $(t-t_*)^{-\alp}$ with
$\alp \ge 1$ and $t_*$ the singularity time. 
In MHD \cite{BKM_MHD}, one deals 
with the Els\"asser fields $\bzpm=\bv\pm \bb$ and
$\bompm=\bom \pm \bj=\nabla \times (\bv \pm \bb)$, with $\bom$ the 
vorticity, $\bv$ the velocity, $\bj$ the 
current density and $\bb=\nabla \times \bA$ the induction in dimensionless Alfvenic units, $\bA$ being the vector potential. Intense current sheets are known 
to form at either magnetic nulls ($\bb\equiv 0$) or when one or two (but 
not all) components of the magnetic field go to zero or have strong 
gradients. In two dimensional configurations, a vortex quadrupole is 
also associated with these structures. The occurrence of singularities 
in MHD has 
been examined in direct numerical simulations (DNS), with either 
regular \cite{PPS95,kerrB1} or adaptive grids \cite{grauer}, and with 
different initial configurations with no clear-cut conclusions in view of 
the necessity for resolving a large range of scales (see \cite{kerr05} and 
references therein for the Euler case).  Laboratory experiments and DNS 
have also studied the ensuing acceleration of particles in the 
reconnection region (see e.g. \cite{gekelman}).

The early development of small scales in such flows is exponential \cite{syro}
 (in the context of turbulent flows, see \eg \cite{sanmin}),
because of the large-scale gradients of the velocity 
and magnetic fields, assumed given, stretching the vorticity and current. The phase beyond the linear stage,
though, is still unclear. In 2D, numerical simulations with periodic 
boundary conditions show that the late-time evolution of non-dissipative 
MHD flows remains at most exponential \cite{fpsm}, 
a point latter confirmed theoretically \cite{klapper1} by examining the 
structure around hyperbolic nulls, although finite dissipation seems to 
set in \cite{PPS89}.

In 3D, most initial conditions develop
sheets that may render the problem quasi two-dimensional locally;
3D MHD flows display a growth of small scales of an exponential nature, although at later times 
a singular behavior may emerge \cite{kerrB1}. In this light, we address in this paper
the early development of structures in 3D and the ensuing evolution in the presence of dissipation.
\vskip0.03truein

The incompressible MHD equations read:
\begin{eqnarray}
&&{ \partial {\bf v} \over \partial t}+ {\bf v} \cdot \nabla {\bf v} =-{1 \over
\rho \sb{0}}\nabla {\cal P} + {\bf j} \times {\bf b} + \nu \nabla \sp{2} {\bf v}  
\nonumber\\
&&{ \partial {\bf b} \over \partial t}= \nabla \times ( {\bf v}\times
{\bf b})
+\eta \nabla \sp{2} {\bf b} 
\label{MHDt}\end{eqnarray}
together with ${\bf \nabla} \cdot {\bf v} =0 ,  \nabla \cdot {\bf b} =0$;  ${\cal P}$ is the pressure, 
$\rho_0=1$ is the constant density, 
and $\nu$ and $\eta$ are  the kinematic viscosity and magnetic diffusivity.
With $\nu=0,\ \eta=0$, the energy 
$E=\left<v^2+b^2\right>/2$ and cross helicity 
$H_C=\left<{\bf v} \cdot {\bf b}\right>/2$, 
are conserved \cite{notez}, with the magnetic helicity 
$H_M=\left<{\bf A} \cdot {\bf b}\right>$ in 3D.
Defining 
$D_{\pm}/Dt=\partial _t + \bzpm\cdot\nabla$, one can symmetrize Eqs. (\ref{MHDt}) and obtain
 \cite{sanmin}:
\begin{eqnarray}
&&{D_{\mp}\bzpm\over Dt}\  = - \nabla {\cal P} \ \ ,\label{zpm}\\
&&{D_{\mp}\bompm\over Dt} = \bompm \cdot \nabla \bzmp
+ \sum _m \nabla z^{\pm}_m \times \nabla z^{\mp}_m\ \ , \label{opm1}
\end{eqnarray}
omitting dissipation. Note that the first term on the \rhs of (\ref{opm1}) is equal to zero 
in 2D; the second term is absent in the Navier-Stokes case and may account for extra growth 
of the generalized vorticities for conducting 
fluids unless the Els\"asser field gradients are parallel.

To study the development of structures in MHD turbulence, we solve 
numerically Eqs. (\ref{MHDt}) using a pseudospectral method in a three 
dimensional box of side $2\pi$ with periodic boundary conditions. 
All computations are de-aliased, 
using the standard $2/3$ rule. With a minimum wavenumber of $k_{min}=1$ 
corresponding to $L_0=2\pi$, a 
linear resolution of $N$ grid points has a maximum wavenumber $k_{max}=N/3$. 
At all times we have $k_D/k_{max} < 1$, 
where $k_D$ is the dissipation wavenumber evaluated using the Kolmogorov 
spectrum (at early times the resolution condition is less stringent). 
Two different initial conditions are used; Table \ref{table:runs} 
summarizes all the runs.

As the system is evolved, we monitor the small scale development by following 
the dynamical evolution of the extrema 
of the generalized vorticities or of their individual components \cite{note1} 
in the spirit of the BKM criterion.

\begin{table}
\caption{\label{table:runs} Runs with an
Orszag-Tang vortex (OT1-4), or with large-scale ABC flows  
and small-scale random noise with a $k^{-3}$ spectrum (RND1-5);
$N$ is the linear resolution.
}
\begin{ruledtabular}
\begin{tabular}{ccc}
Run         & $N^3$      &$\nu=\eta$ \\
\hline
OT1 - OT4   & $64^3$ -- $512^3$  &$1\times 10^{-2}$ -- $7.5\times 10^{-4}$    \\
RND1 - RND4 & $64^3$ -- $512^3$  &$8\times 10^{-3}$ -- $6\times 10^{-4}$      \\
RND5        & $1536^3$     &$2\times 10^{-4}$     \\
\end{tabular} \end{ruledtabular} \end{table}

\begin{figure}
\centerline{\includegraphics[width=8.4cm]{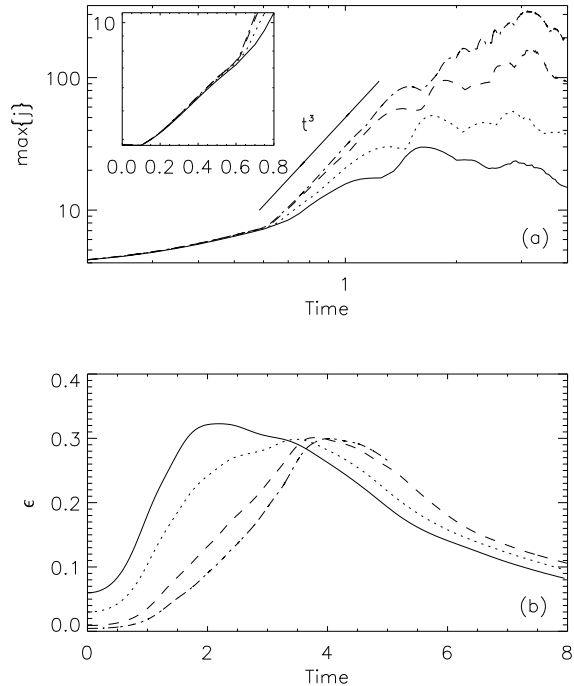}}
\caption{(a) Evolution of the supremum of current for the 
OT runs in log-log. The inset shows the evolution at early times in 
lin-log units; a slope of 
$t^3$ is also indicated. The exponential phase ends at $t\sim 0.6$.  (b) Total dissipation as a function of time 
for the same runs in lin-lin. $R_e = 570$ (solid), $R_e = 1040$ (dot), 
$R_e = 3040$ 
(dash), and $R_e = 5600$ (dash-dot).}
\label{fig_OTdissrate}
\end{figure}

We start discussing the results for the Orszag-Tang vortex (OT hereafter); in two dimensions \cite{OT2d}, it has become a prototype flow for the study of MHD turbulence, including in the compressible case \cite{dahl1}. 
In 3D, the velocity and the magnetic field are taken to be:
\begin{eqnarray}
&& \bv_{0}=[-2 \sin y,2 \sin x, 0] \ \ ,  \nonumber \\
&& \bb_{0}=\beta[-2 \sin 2y+ \sin z, 2\sin x + \sin z,\sin x +\sin y] .
\nonumber \end{eqnarray}

The OT flow in 2D has a stagnation point in the $(x,y)$ plane 
and an hyperbolic X-point for the magnetic field; a 
3D perturbation is added in the $z$ direction, 
resulting in a flow that has nulls for the magnetic field (three components 
equal to zero) of different types \cite{PF00}
corresponding to the signs of the eigenvalues of the $\partial_i b_j$ matrix \cite{PPS95}; 
initially, the kinetic and magnetic 
energy $E_V$ and $E_M$ are equal to 2 with $\beta =0.8$, 
 the normalized 
correlation 
$\tilde \rho^c=2\left<\bv \cdot \bb\right>/(\left<v^2+b^2\right>) \sim 0.41$, and $H_M=0$.

Four runs were done for spatial resolutions up to 
$512^3$. The Reynolds number $R_e=UL/\nu$ (where $U=\left<v^2\right>$ is 
the rms velocity, $L=2\pi \int E_V(k)k^{-1} dk/\int E_V(k)dk$ is the integral 
scale, and $E_V(k)$ is the kinetic energy spectrum) ranges from $570$ to 
$5600$ at the time of maximum dissipation of energy 
$\epsilon = -\nu \left< \omega^2 \right> -\eta \left< j^2 \right>$.
Figure \ref{fig_OTdissrate}(a) shows the temporal evolution of the 
maximum of the current $\max\{j\}$ (the vorticities $\bom$ and $\bompm$ 
behave in a similar fashion). 
After an initial exponential phase up to $t\sim 0.6$ and corresponding 
to the linear development of current (and vorticity) sheets through 
stretching by velocity gradients, a faster increase develops with, at 
high resolution, a self-similar $\sim t^3$ law. Note that the 
growth of $\max\{j\}$ during the early exponential phase seems to be 
independent of the value of $R_e$.

The first temporal maximum of $\max\{j\}$ is reached at slightly later times 
as $R_e$ increases; similarly [see 
Fig. \ref{fig_OTdissrate}(b)], 
the total energy dissipation 
$\epsilon$ shows a delay in the 
onset of the development of small scales as $R_e$ grows, 
reminiscent of 2D behavior \cite{PPS89},
with a slower global growth rate after the initial exponential 
phase; this delay, however, does not preclude reaching a quasi-constant 
maximum of $\epsilon$ in time as $R_e$ grows. Whereas in the 2D case, the 
constancy of $\epsilon$ only obtains at later times when reconnection 
sets in, with a multiplicity of current sheets, in the 3D case more 
instabilities of current and vorticity structures are possible 
and the flow becomes complex as soon as the linear phase has ended. 
The dependence on $R_e$ of the time at which the first 
maximum of $\max\{j\}$ is reached  is slow ($\sim R_e^{0.08}$), and similarly for the time the maximum of 
$\epsilon$ is reached.
Computations at higher $R_e$ should be performed
to confirm these results.

\begin{figure}
\centerline{\includegraphics[width=8.4cm]{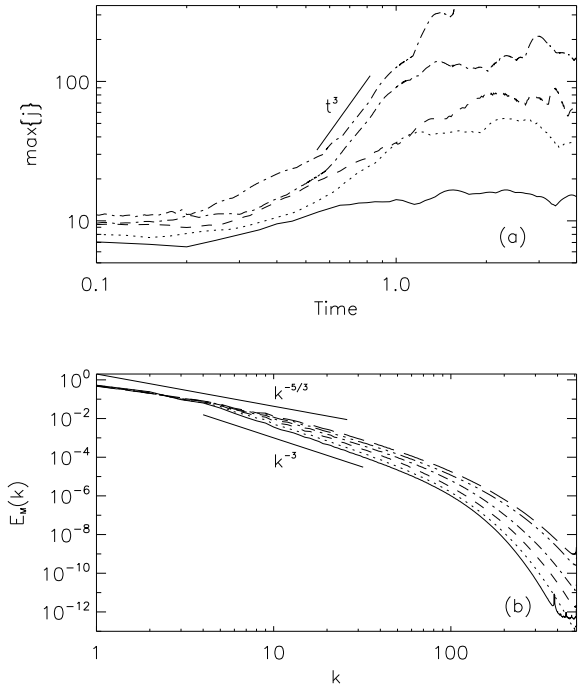}}
\caption{(a) Evolution of the supremum of the current density in 
log-log for runs RND1 to RND5, with $R_e=690$ (solid), $R_e=1300$ (dot), 
$R_e=2300$ (dash), $R_e=4200$ (dot-dash), and $R_e=10100$ (long dash). 
At high $R_e$, a power law consistent with $t^3$ is recovered. (b) Magnetic 
energy spectra at early times in run RND5. The lines (from below) 
correspond to $t=0.6$ up to $t=1.6$ with temporal increments of $0.2$. Slopes 
of $k^{-3}$ and $k^{-5/3}$ are indicated as a reference.}
\label{fig_RNDllogmaxc}
\end{figure}

The sharp transition around $t=0.6$ can be interpreted in terms of the 
non-locality of interactions in MHD turbulence \cite{alex_mhd} with transfer of energy involving widely 
separated scales. 
Thus, as the flow exits the linear phase, all scales can interact 
simultaneously; this may be a reason why, in a computation using OT and 
adaptive mesh refinement \cite{grauer}, it was found to be difficult 
to go beyond $t=0.6$ since small scales were developing abruptly in many 
places in the flow through prevailing nonlinearities. The energy spectra 
in this early phase are
steep, with a power law $\sim k^{-3}$ (not shown). A shallower $\sim k^{-1.70}$ spectrum develops
at later times, as found in earlier works.

In view of similarities between the behavior observed on the 
3D OT vortex and its 2D counter-part, it is worth asking whether such 
a development is not due to the high degree of symmetry of the flow. In 
that light, we now examine the temporal development of a Beltrami flow 
on which small scale random fluctuations are added. 
The initial velocity and magnetic 
field spectra  are taken $\sim k^{-3} e^{-2{(k/k_0)}^2}$; the shells 
with $k\in [1,3]$ have a superposition of three ABC flows
\cite{ABC}, and the rest of the spectrum is loaded with Fourier 
modes with Gaussian random realizations for their phases chosen so that initially,
 $E_V=E_M=0.5$, ${\tilde {\rho}}^c\sim 10^{-4}$ and $H_M\sim 0.45$. 
Unlike the OT runs, there are no 2D null 
points or exact zeros in the magnetic field. As for the OT case, four 
runs (RND1-RND4) were done with resolutions ranging from $64^3$ to 
$512^3$ grid points; RND5 on a grid of
$1536^3$ points is run until saturation of growth of the maximum current.

Both the exponential and the self-similar phases are noisier (see Fig. 
\ref{fig_RNDllogmaxc}a), as can be expected with several structures 
competing for the small scale development of maxima. At low $R_e$, 
self-similar evolution seems to occur at a slower pace, with laws 
$\sim t^2$, as in fact also found in 2D at comparable
resolutions. However, the two runs with highest 
resolution (RND4 and RND5) indicate a steeper development compatible 
with a $t^3$ law.

Figure \ref{fig_RNDllogmaxc}(b) shows the evolution of the magnetic 
energy spectrum $E_M(k)$ at early times, during the self-similar growth of the current 
density [the evolution of $E_V(k)$ is similar]. Before 
$t \sim 0.6$, the largest wavenumbers have amplitudes 
 of the order of the truncation error. For $t \ge 0.6$, as all scales are nonlinearly
excited, a self-similar growth sets in and the 
energy spectra are compatible with a $k^{-3}$ law. After $\max\{j\}$ 
saturates, the slope of $E_M(k)$ increases slowly towards a $k^{-1.70}$ 
law. The same behavior is observed in the OT run, in which no 
$k^{-3}$ power law is imposed in the initial conditions.

\begin{figure}
\centerline{\includegraphics[width=8.5cm]{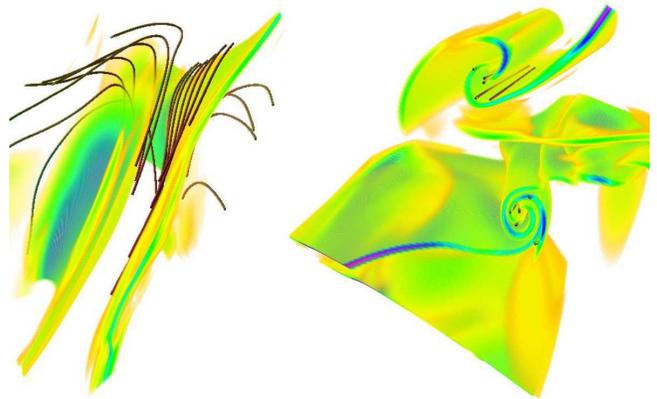}}
\caption{(Color online) Regions of strong current density, and magnetic 
field lines in their vicinity for run RND5 at $t=1.6$. The region at left has $450^2 \times 250$ points, and that at right has $260 \times 160 \times 200$. 
The sheets are thin and elongated (up to 1/3 the size of the box); the magnetic field lines are parallel to the sheet and quasi-orthogonal to each other on each side of it, and they depart from the sheet transversally. Both folding (left) and rolling (right) occurs at this $R_e$. Vortex sheets (not shown) are 
co-located and parallel to the current sheets.}
\label{fig_current2.jpg}
\end{figure}

The structures that develop appear to be accumulations of current sheets (similarly 
for the vorticity, not shown), as was already found in \cite{PPS95}. Figure 
\ref{fig_current2.jpg} shows a zoom on two such structures, with the magnetic 
field lines indicated as well. It appears clear from such figures that 
only one component of the magnetic field reverses sign in most of these 
configurations, reminiscent of magnetospheric observations.
Both terms appearing in Eq. (\ref{opm1}) for the dynamical evolution of 
$\bompm$ are substantial and comparable in magnitude although they may be quite weak elsewhere in the flow. 
Kelvin-Helmoltz instabilities with rolling up of such 
sheets are also present in the flow but only at the highest Reynolds number (run RND5); at lower $R_e$ the sheets are thicker, the instability is too slow and only folding of such sheets occur.
Magnetic field lines are parallel to 
the roll, in such a way that magnetic tension does not prevent the occurrence 
of the instability. Note that folding of magnetic 
structures has been advocated in the context of 
MHD at large magnetic Prandtl number \cite{cowley}. Alfvenization of the flow 
($\bv = \pm \bb$) is rather strong in the vicinity of the sheets, with $0.7\le |{\tilde{\rho}}^c| \le 1$, although globally the flow remains uncorrelated 
(${\tilde{\rho}}^c\sim 4 \times 10^{-4}$);  this local Alfvenization gives stability to such structures since the nonlinear terms are weakened, in much the same way vortex tubes in Navier-Stokes flows are long-lived because of (partial) Beltramization ($\bv \sim \pm \bom$). Moreover, within the sheet ${\tilde{\rho}}^c$ is positive, and it is negative outside, with a slight predominance of $\bb$. All this indicates that a double velocity-magnetic field shear plays an important role in the development of small scales in MHD.

There is an elementary, analytically-soluble, one-dimensional model that 
illustrates sharply the role that velocity shear can play in enhancing 
current intensity, \eg during early dynamo activity \cite{montgo}. 
This consists of two semi-infinite slabs of rigid metal with equal 
conductivities, at rest against each other at the plane $y=0$, say. A 
uniform dc magnetic field ${\bf b}_0$ is perpendicular to the interface 
and penetrates both slabs. At time $t=0$, the slabs are impulsively 
forced to slide past each other in the $x$-direction with equal and 
opposite velocities (${\bf v}_0$, say). The developing (quasi-magnetostatic) 
field, which acquires an $x$-component, is a function of $y$ and $t$ only, 
and is governed by diffusion equations above and below the plane $y=0$. 
Matching tangential components of the electric field immediately above 
and below the interface reduces the pair to a soluble diffusion equation 
with a fixed $y$-derivative at $y=0$. The resulting magnetic field is 
expressible in terms of complementary error functions and grows without 
bound, as does the total Ohmic dissipation. The introduction of 
a time dependence in ${\bf v}_0$ may allow also for solutions 
in which the maximum of the current grows as a power law in time.

In conclusion, high resolution simulations of the early stages in the development of 
MHD turbulence allowed us to study the creation and evolution of small scale 
structures in three-dimensional flows. Roll up of current and vortex sheets, 
and a self-similar growth of current and vorticity maxima was found, features that 
to the best of our knowledge were not observed in previous simulations at 
smaller Reynolds numbers. Also, a convergence of the maximum dissipation 
rate to a value independent of $R_e$ was found. More analysis will be carried out
to understand how structures are formed, the relevance of the development of 
alignment between the fields and the creation and role of local exact solutions to the 
MHD equations (such as Alfv\'en waves).

\vskip0.15truein

{\it 
NSF grants CMG-0327888 and ATM-0327533 are acknowledged. Computer 
time provided by NCAR. Three-dimensional visualizations were done using VAPoR.
\cite{vapor}.}


\begin{thebibliography}{1}
\bibitem{note1}
One may also monitor the temporal development of the symmetrized velocity 
and magnetic field gradient matrices, as well as that of the total 
enstrophy production \cite{kerrB1}.
\end{thebibliography}
\end{document}